\documentstyle[12pt]{article}
\newcommand{\tr}[1]{\,{\rm tr}\,#1\,}

\begin{document}
\def\newmathop#1{\mathop{#1}\limits}
\def\osim#1{\displaystyle\newmathop{#1}_{\lambda\to 0}}
\def\linfty#1{\displaystyle\newmathop{#1}_{s\to \infty}}
\def\rr#1{\displaystyle\newmathop{#1}_{~~s \to \infty ,~ t~fixed~~}}
\title{
\begin{flushright}
{\small CVV-201-95 \\
SMI-05-95 \\ HEP-TH  9502092}
\end{flushright}
\vspace{2cm}
 The Master Field for Rainbow Diagrams
\\ and \\ Free Non-Commutative Random Variables }
\author{L. Accardi,$~~~$
I.Ya.Aref'eva \thanks{Steklov Mathematical Institute,
Vavilov 42, GSP-1, 117966, Moscow, e-mail: Arefeva@arevol.mian.su}$~~~$
and $~~$ I.V.Volovich \thanks{On leave absence from
Steklov Mathematical Institute,
Vavilov 42, GSP-1, 117966, Moscow}
\\$~$\\
{\it Centro V. Volterra Universit\`a di Roma Tor Vergata}\\
{\it via di Tor Vergata, 00133 -- Roma}\\
}
\date {$~$}
\maketitle
\begin {abstract}
The master field for a subclass of planar diagrams, so called rainbow
diagrams, for higher dimensional large N theories is  considered.
An explicit representation for the master field in terms of
noncommutative random variables in the modified interaction representation
in the Boltzmannian Fock space is given. A natural interaction in the
Boltzmannian Fock space is formulated by means of a rational
function of the interaction Lagrangian instead of the ordinary
exponential function in the standard Fock space.
\end {abstract}
\newpage
\section{Introduction}
\setcounter{equation}{0}
The problem of summation of all planar diagrams in higher dimensional
space-time  is still out of reach.
Its solution is closely related with problem of finding the leading
asymptotics in  matrix models for large N and may have important
applications to the hadron dynamics  \cite {tH,Ven,Wit}.
Summation of planar diagrams has been performed only
in low dimensional space-time \cite {BIPZ}.

One can write a closed system of equations for invariant correlation
functions in the large N limit, so called the planar Schwinger-Dyson
equations, for arbitrary dimension of space-time.
In the early 80-s it was suggested  \cite {Wit} that
there exists the master field $\Phi (x)$ such that the
correlation functions for this  field $\Phi (x)$ are equal to
the large N limit of  invariant correlation functions
for matrix models,
\begin {equation} 
                                                          \label {mc}
\lim _{N\to\infty} \frac{1}{N^{1+n/2}}<\tr (M(x_{n})...M(x_{1}))>
= <0|\Phi (x_{n}) ...\Phi (x_{1})|0>
\end   {equation} 

It was suggested  that $\Phi (x)$ satisfies the equation \cite {Haan}
\begin {equation} 
                                                          \label {omf}
[i \frac{\delta S[\Phi ]}{\delta \Phi (x)}+2\Pi (x)]|0>=0
\end   {equation} 
where $S$ is an action and  $\Pi $ and $\Phi $ are the subjects of
the relation \cite {Haan,HalSc}
\begin {equation} 
                                                          \label {pcr}
[\Pi (x),\Phi (y)] =i\delta ^{(D)}(x-y)|0><0|
\end   {equation} 
An operator realization of this algebra
proposed in  \cite {Haan} has used the knowledge of all correlation
functions. This was considered as an evident drawback of such approach.
In that time it was also discussed the problem of
finding a generating functional reproducing the planar Schwinger-Dyson
equations or equations (\ref {omf}), (\ref {pcr}). It was
pointed out that the  generating functional cannot depend on one
commutative source \cite {Cvet}. It was proposed to use some
auxiliary fermionic fields to reproduce the planar Schwinger-Dyson
equations \cite {Ar}. For gauge theories the suitable
generating functional  is nothing but a functional on paths,
i.e. the Wilson loops, that satisfies the Makeenko-Migdal equation \cite {MM}.
The stochastic equation for large N master fields was proposed in \cite {GH}.

Recently it has been a reveal of an interest to the problem of constructing
the master field for planar graphs. One of origins for this are mathematical
works \cite {Voi,Voi91,Sin}  devoted to non-commutative probability, for
a review see \cite {Ac,ALV}.  Singer has advocated
that the essential difficulty of the large number of degrees of freedom in
higher dimensional large N  matrix models is dealt with finding master
fields which live in "large" operator algebras such as the type
$II_{1}$ factor associated with free group. In the recent paper by Gopakumar
and Gross  \cite {GG} the basic concepts of non-commutative probability
have been reviewed and applied to the large N limit of matrix models.
They stress that if one can solve a matrix model then one can write an explicit
expression for the master field as an operator in a well defined Hilbert space.
Douglas also proposed to use ideas of non-commutative probability
to the large N stochastic approach  \cite {Doug3}.
The explicit construction of the master fields for several
low-dimensional models including
$QCD_{2}$ has been given  \cite {GG}-\cite {DL}.

However an {\it effective } (i.e. without the knowledge of
correlation fonctions) operator realization for the master field
for all planar diagrams in higher dimensional space-time is still unknown.
Therefore it is worth to try to find an effective operator realization of
the master field  for some subset of the planar diagrams.
The goal of this letter is to construct an explicit operator realization of
the master field   for {\it rainbow} graphs. Rainbow graphs form a subset of
planar graphs. It is turn out that rainbow correlation functions
may be obtained by average of the fields with Boltzman statistics.
The construction does not require as input correlation functions.
More exactly we show that to get a closed set of equations
for correlations functions for a model with an interaction
in the Boltzmannian Fock space one has to deal with a
modified interaction representation. This new interaction representation
involves not the ordinary exponential function  of the
interaction but a rational function and will be given by the formula
\begin {equation} 
                                                          \label {nH}
<\phi (x_{m})...\phi (x_{1}){1\over {1-g\int dy
V_{int}(\phi (y))}}>
\end   {equation} 

The paper is organized as follows. In Sect.2 we present the
Schwinger-Dyson equations for  rainbow  diagrams.
Sect.3 contains a necessary information about the Boltzmannian
Fock space. We argue also that to get a closed set of the
Schwinger-Dyson type of equation we have to deal with
the interaction representation in the Boltzmannian Fock space
in the form (\ref {nH}). We also show that the corresponding
Schwinger-Dyson equations reproduce the
Schwinger-Dyson equations for  rainbow  diagrams.

\section{Rainbow Diagrams}
\setcounter{equation}{0}
Let us consider the correlation functions of the form
\begin {equation} 
                                                          \label {gv}
<{\cal V}(x_{n},...x_{1})>=\frac{1}{N^{1+n/2}}<\tr (M(x_{n})...M(x_{1}))>
\end   {equation} 
$<\cdot>$ means
\begin {equation} 
                                                          \label {<>}
<{\cal O}(M)>=\frac{1}{Z}\int {\cal O}(M)\exp \{-S[M]\}dM,
\end   {equation} 
where
\begin {equation} 
                                                          \label {ac}
S[M]=\int dx
[\frac{1}{2}\tr (M(-\bigtriangleup  +m^{2})M)
+\frac{g}{4N}\tr M^{4}(x)]
\end   {equation} 
\unitlength=0.7mm
\linethickness{0.4pt}
 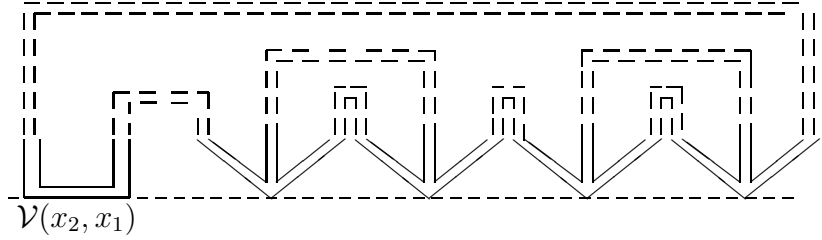
\begin{figure}
\begin{center}
\begin{picture}(140.00,40.00)
\put(55.00,5.00){\line(5,4){14.07}}
\put(66.00,16.00){\line(-5,-4){10.00}}
\put(56.00,8.00){\line(0,1){11.07}}
\put(54.00,19.00){\line(0,-1){11.00}}
\put(54.00,8.00){\line(-4,3){10.93}}
\put(41.00,16.00){\line(5,-4){14.07}}
\put(85.00,5.00){\line(5,4){14.07}}
\put(96.00,16.00){\line(-5,-4){10.00}}
\put(86.00,8.00){\line(0,1){11.07}}
\put(84.00,19.00){\line(0,-1){11.00}}
\put(84.00,8.00){\line(-4,3){10.93}}
\put(71.00,16.00){\line(5,-4){14.07}}
\put(115.00,5.00){\line(5,4){14.07}}
\put(126.00,16.00){\line(-5,-4){10.00}}
\put(116.00,8.00){\line(0,1){11.07}}
\put(114.00,19.00){\line(0,-1){11.00}}
\put(114.00,8.00){\line(-4,3){10.93}}
\put(101.00,16.00){\line(5,-4){14.07}}
\put(145.00,5.00){\line(5,4){14.07}}
\put(156.00,16.00){\line(-5,-4){10.00}}
\put(146.00,8.00){\line(0,1){11.07}}
\put(144.00,19.00){\line(0,-1){11.00}}
\put(144.00,8.00){\line(-4,3){10.93}}
\put(131.00,16.00){\line(5,-4){14.07}}
\put(8.00,16.00){\line(0,-1){10.93}}
\put(8.00,5.07){\line(1,0){20.00}}
\put(25.07,7.07){\line(-1,0){14.00}}
\put(28.00,5.07){\line(0,1){10.93}}
\put(25.00,16.00){\line(0,-1){9.00}}
\put(11.07,7.07){\line(0,1){8.93}}
\put(18.00,1.00){\makebox(0,0)[cc]{${\cal V}(x_{2},x_{1})$}}
\put(8.00,17.00){\line(0,1){3.00}}
\put(8.00,22.00){\line(0,1){3.00}}
\put(8.00,27.00){\line(0,1){3.00}}
\put(8.00,32.00){\line(0,1){3.07}}
\put(8.00,37.00){\line(0,1){3.00}}
\put(10.00,17.00){\line(0,1){3.00}}
\put(10.00,22.00){\line(0,1){3.00}}
\put(10.00,27.00){\line(0,1){3.00}}
\put(10.00,32.00){\line(0,1){3.00}}
\put(10.00,37.00){\line(0,1){3.00}}
\put(25.00,17.00){\line(0,1){3.00}}
\put(25.00,22.00){\line(0,1){3.00}}
\put(28.00,17.00){\line(0,1){2.00}}
\put(28.00,21.00){\line(0,1){2.00}}
\put(41.00,17.00){\line(0,1){3.00}}
\put(41.00,19.00){\line(0,1){1.00}}
\put(43.00,25.00){\line(0,-1){3.00}}
\put(43.00,20.00){\line(0,-1){3.00}}
\put(54.00,21.00){\line(0,1){3.00}}
\put(54.00,29.00){\line(0,-1){3.00}}
\put(54.00,31.00){\line(0,1){2.00}}
\put(54.00,33.07){\line(1,0){3.00}}
\put(56.00,21.00){\line(0,1){3.00}}
\put(56.00,26.00){\line(0,1){3.00}}
\put(56.00,29.07){\line(0,0){0.00}}
\put(56.00,29.07){\line(0,-1){0.00}}
\put(67.00,17.00){\line(0,1){3.00}}
\put(67.00,22.00){\line(0,1){3.00}}
\put(69.00,17.00){\line(0,1){3.00}}
\put(69.00,22.00){\line(0,1){2.00}}
\put(71.00,17.00){\line(0,1){3.00}}
\put(71.00,22.00){\line(0,1){2.00}}
\put(73.00,17.00){\line(0,1){3.00}}
\put(73.00,22.00){\line(0,1){3.00}}
\put(84.00,21.00){\line(0,1){3.00}}
\put(84.00,26.00){\line(0,1){3.00}}
\put(86.00,21.00){\line(0,1){3.00}}
\put(86.00,26.00){\line(0,1){3.00}}
\put(86.00,31.00){\line(0,1){2.00}}
\put(97.00,17.00){\line(0,1){3.00}}
\put(97.00,21.00){\line(0,1){3.00}}
\put(99.00,17.00){\line(0,1){3.00}}
\put(99.00,22.00){\line(0,1){2.00}}
\put(101.00,17.00){\line(0,1){3.00}}
\put(101.00,22.00){\line(0,1){2.00}}
\put(103.00,16.00){\line(0,1){3.00}}
\put(103.00,21.00){\line(0,1){3.00}}
\put(114.00,33.00){\line(0,-1){3.00}}
\put(114.00,29.00){\line(0,-1){3.00}}
\put(114.00,24.00){\line(0,-1){3.00}}
\put(116.00,21.00){\line(0,1){3.00}}
\put(116.00,26.00){\line(0,1){3.00}}
\put(127.00,17.00){\line(0,1){3.00}}
\put(127.00,22.00){\line(0,1){3.00}}
\put(129.00,24.00){\line(0,-1){2.00}}
\put(129.00,20.00){\line(0,-1){3.00}}
\put(131.00,17.00){\line(0,1){3.00}}
\put(131.00,22.00){\line(0,1){2.00}}
\put(133.00,26.00){\line(0,-1){3.00}}
\put(133.00,21.00){\line(0,-1){3.00}}
\put(144.00,29.00){\line(0,-1){3.00}}
\put(144.00,24.00){\line(0,-1){3.00}}
\put(146.00,21.00){\line(0,1){3.00}}
\put(146.00,26.00){\line(0,1){3.00}}
\put(146.00,29.00){\line(0,1){2.00}}
\put(146.00,31.00){\line(0,1){2.00}}
\put(25.00,25.00){\line(1,0){3.00}}
\put(30.00,25.00){\line(1,0){3.00}}
\put(36.00,25.00){\line(1,0){3.00}}
\put(33.07,23.07){\line(-1,0){3.00}}
\put(39.00,23.00){\line(-1,0){3.00}}
\put(41.00,25.00){\line(1,0){2.00}}
\put(56.00,31.00){\line(1,0){3.00}}
\put(59.00,33.00){\line(1,0){3.00}}
\put(61.00,31.00){\line(1,0){3.00}}
\put(62.00,33.00){\line(-1,0){0.95}}
\put(62.00,33.00){\line(-1,0){0.95}}
\put(66.00,33.00){\line(1,0){2.95}}
\put(66.00,31.00){\line(1,0){2.95}}
\put(67.00,26.00){\line(1,0){1.95}}
\put(71.00,26.00){\line(1,0){1.95}}
\put(69.00,24.00){\line(1,0){2.00}}
\put(71.00,31.00){\line(1,0){2.98}}
\put(73.00,26.00){\line(-1,0){3.00}}
\put(73.00,26.00){\line(-1,0){3.00}}
\put(76.00,31.00){\line(1,0){3.00}}
\put(76.00,33.00){\line(-1,0){3.00}}
\put(81.00,33.00){\line(-1,0){3.00}}
\put(81.00,31.00){\line(1,0){3.00}}
\put(85.99,32.95){\line(-1,0){2.97}}
\put(97.00,26.00){\line(1,0){2.00}}
\put(99.00,24.00){\line(1,0){2.00}}
\put(103.00,26.00){\line(-1,0){2.03}}
\put(114.00,33.00){\line(1,0){2.00}}
\put(116.00,31.00){\line(1,0){3.00}}
\put(121.00,33.00){\line(-1,0){3.00}}
\put(121.00,31.00){\line(1,0){3.00}}
\put(126.00,31.00){\line(1,0){3.00}}
\put(127.00,26.00){\line(1,0){2.00}}
\put(129.00,24.00){\line(1,0){2.00}}
\put(130.00,26.00){\line(1,0){3.00}}
\put(131.00,26.00){\line(1,0){2.00}}
\put(131.00,26.00){\line(1,0){2.00}}
\put(131.00,26.00){\line(-1,0){1.00}}
\put(131.00,31.00){\line(1,0){3.00}}
\put(136.00,31.00){\line(1,0){3.00}}
\put(141.00,31.00){\line(1,0){3.00}}
\put(126.00,33.00){\line(-1,0){3.00}}
\put(131.00,33.00){\line(-1,0){3.00}}
\put(136.00,33.00){\line(-1,0){3.00}}
\put(141.00,33.00){\line(-1,0){3.00}}
\put(146.00,33.00){\line(-1,0){3.00}}
\multiput(5.00,5.00)(3.5,0){43}{\line(1,0){2.00}}
\multiput(10.00,40.00)(4.5,0){32}{\line(1,0){3.00}}
\multiput(8.00,42.00)(4.5,0){33}{\line(1,0){3.00}}
\multiput(156.00,17.00)(0,5){5}{\line(0,1){3.00}}
\multiput(158.00,17.00)(0,5){5}{\line(0,1){3.00}}
\end{picture}
\end{center}
\caption{Wick theorem for rainbow diagrams}\label{f1}
\end{figure}

Here $M(x)$ is $N\times N$ matrix function.
We assume all necessary regularizations. For our purpose
it is essential that the regularization is such that the
free propagator is
\begin {equation} 
                                                          \label {pr}
<M_{ij}(x)M_{j'i'}(y)>^{(0)}=\delta _{ii'}\delta _{jj'} D(x-y),
\end   {equation} 
\begin {equation} 
                                                          \label {fpr}
(-\bigtriangleup  +m^{2})_{x}D(x-y)=\delta ^{(D)}(x-y).
\end   {equation} 

We shall consider the external lines corresponding to global invariant Green
 functions as the lines corresponding to generalized vertex.
The rainbow diagrams for$~~$  $<{\cal V}(x_{n},...x_{1})>$ are defined
as a part of planar non-vacuum diagrams which are topologically equivalent
to the graphs with all vertexes lying on some
stright line on the the right of generalized vertex  and all propagators lying
in the
half plane. The rainbow diagramms are illustraited for the $M^3$-
interaction on Fig.\ref {f1}, where vertexies are drawn by solid double
lines and all contraction (propagators) by double dash lines;
in that follows we will use
also solid lines for propagators.

To write down the rainbow Schwinger-Dyson equations one has to select
rainbow diagrams from the both hand sides of the planar Schwinger-Dyson
equations.
In the large N limit  the planar Schwinger-Dyson equations have
 the form
\begin {equation} 
                                                          \label {pl}
(-\bigtriangleup  +m^{2})_{x_{l}}G_{n}(x_{n},...x_{1})=
gG_{n+2}(x_{n},...x_{l+1},x_{l},x_{l},x_{l},x_{l-1},...x_{1})
\end   {equation} 
$$
+\sum _{i< l}\delta (x_{l}-x_{i})G_{l-i-1}(x_{i-1},...x_{l+1})
G_{n+i-l-1}(x_{n},...x_{l+1},x_{i-1},...x_{1})$$
$$
+\sum _{l< i}\delta (x_{l}-x_{i})G_{i-l-1}(x_{l-1},...x_{i+1})
G_{n+l-i-1}(x_{n},...x_{i+1},x_{l-1},...x_{1}),$$
where
\begin {equation} 
                                                          \label {G}
G_{n}(x_{n},...x_{1})=
\lim _{N\to\infty} \frac{1}{N^{1+n/2}}<\tr (M(x_{n})...M(x_{1}))>
\end   {equation} 
The planar Schwinger-Dyson equations (\ref {pl}) are written for
the case of quartic interaction (\ref {ac})
and they are symbolically presented on
Fig.\ref {fig2}.
\unitlength=1mm
\special{em:linewidth 0.4pt}
\linethickness{0.4pt}
    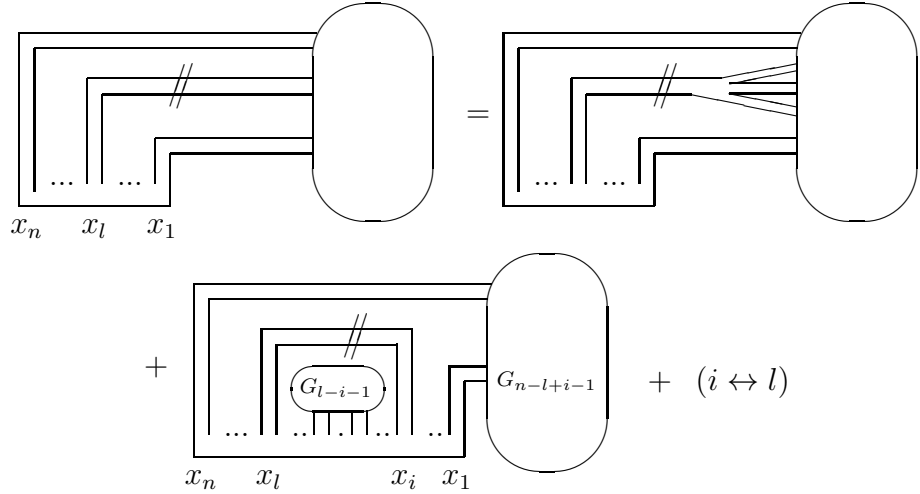
\begin{figure}
\begin{center}
\begin{picture}(63.00,35.00)
\put(5.00,5.00){\line(0,1){23.00}}
\put(5.07,28.00){\line(1,0){38.93}}
\put(7.00,26.00){\line(0,-1){19.00}}
\put(9.00,8.00){\makebox(0,0)[lc]{$...$}}
\put(7.00,8.00){\line(0,-1){0.07}}
\put(5.00,5.00){\line(1,0){20.07}}
\put(25.07,5.07){\line(0,1){6.93}}
\put(25.07,12.00){\line(1,0){18.93}}
\put(44.00,12.00){\line(0,1){2.00}}
\put(44.00,14.00){\line(-1,0){20.93}}
\put(23.07,14.00){\line(0,-1){6.00}}
\put(14.00,8.00){\line(0,1){14.00}}
\put(14.00,22.00){\line(1,0){20.00}}
\put(44.00,18.00){\line(0,0){0.00}}
\put(30.00,19.87){\line(-1,0){14.00}}
\put(16.00,19.87){\line(0,-1){11.87}}
\put(18.00,8.00){\makebox(0,0)[lc]{$...$}}
\put(52.00,17.50){\oval(16.00,29.00)[]}
\put(25.07,18.27){\line(1,3){1.87}}
\put(28.00,24.00){\line(-1,-3){2.00}}
\put(7.00,26.00){\line(1,0){37.00}}
\put(44.53,28.00){\line(-1,0){6.53}}
\put(5.00,5.00){\line(0,1){0.07}}
\put(7.00,26.00){\line(0,-1){0.07}}
\put(32.00,22.00){\line(1,0){12.00}}
\put(44.00,19.87){\line(-1,0){14.80}}
\put(66.00,17.00){\makebox(0,0)[cc]{$=$}}
\put(24.00,2.00){\makebox(0,0)[cc]{$x_{1}$}}
\put(6.00,2.00){\makebox(0,0)[cc]{$x_{n}$}}
\put(15.00,2.00){\makebox(0,0)[cc]{$x_{l}$}}
\end{picture}
\begin{picture}(60.00,35.00)
\put(5.00,5.00){\line(0,1){23.00}}
\put(5.07,28.00){\line(1,0){38.93}}
\put(7.00,26.00){\line(0,-1){19.00}}
\put(9.00,8.00){\makebox(0,0)[lc]{$...$}}
\put(7.00,8.00){\line(0,-1){0.07}}
\put(5.00,5.00){\line(1,0){20.07}}
\put(25.07,5.07){\line(0,1){6.93}}
\put(25.07,12.00){\line(1,0){18.93}}
\put(44.00,12.00){\line(0,1){2.00}}
\put(44.00,14.00){\line(-1,0){20.93}}
\put(23.07,14.00){\line(0,-1){6.00}}
\put(14.00,8.00){\line(0,1){14.00}}
\put(14.00,22.00){\line(1,0){20.00}}
\put(34.00,22.00){\line(5,1){10.00}}
\put(44.00,23.00){\line(-5,-1){8.93}}
\put(35.07,21.33){\line(1,0){8.93}}
\put(44.00,20.00){\line(-1,0){8.93}}
\put(44.00,18.00){\line(0,0){0.00}}
\put(35.00,20.00){\line(5,-1){9.00}}
\put(44.00,17.00){\line(-5,1){14.00}}
\put(30.00,19.87){\line(-1,0){14.00}}
\put(16.00,19.87){\line(0,-1){11.87}}
\put(18.00,8.00){\makebox(0,0)[lc]{$...$}}
\put(52.00,17.50){\oval(16.00,29.00)[]}
\put(25.07,18.27){\line(1,3){1.87}}
\put(28.00,24.00){\line(-1,-3){2.00}}
\put(7.00,26.00){\line(1,0){37.00}}
\put(44.53,28.00){\line(-1,0){6.53}}
\end{picture}
\begin{picture}(80.00,35.00)
\put(1.00,19.00){\makebox(0,0)[cc]{$+$}}
\put(6.07,30.00){\line(1,0){38.93}}
\put(10.00,10.00){\makebox(0,0)[lc]{$...$}}
\put(6.00,7.00){\line(1,0){20.07}}
\put(45.00,14.00){\line(0,1){2.00}}
\put(15.00,10.00){\line(0,1){14.00}}
\put(15.00,24.00){\line(1,0){20.00}}
\put(17.00,21.87){\line(0,-1){11.87}}
\put(19.00,10.00){\makebox(0,0)[lc]{$..$}}
\put(53.00,19.50){\oval(16.00,29.00)[]}
\put(26.07,20.27){\line(1,3){1.87}}
\put(29.00,26.00){\line(-1,-3){2.00}}
\put(8.00,28.00){\line(1,0){37.00}}
\put(45.53,30.00){\line(-1,0){6.53}}
\put(29.07,21.87){\line(1,0){4.00}}
\put(37.00,10.00){\makebox(0,0)[lc]{$..$}}
\put(42.00,7.00){\line(0,1){10.07}}
\put(42.00,17.07){\line(1,0){3.07}}
\put(45.00,19.00){\line(-1,0){5.00}}
\put(40.00,19.07){\line(0,-1){9.07}}
\put(42.00,7.00){\line(-1,0){16.00}}
\put(6.00,7.00){\line(0,1){23.00}}
\put(8.00,28.00){\line(0,-1){18.00}}
\put(22.00,10.00){\line(0,1){3.07}}
\put(24.00,13.00){\line(0,-1){3.00}}
\put(25.00,10.00){\makebox(0,0)[lc]{$.$}}
\put(30.00,10.00){\makebox(0,0)[lc]{$..$}}
\put(25.00,16.00){\makebox(0,0)[cc]{$_{G_{l-i-1}}$}}
\put(7.00,4.00){\makebox(0,0)[cc]{$x_{n}$}}
\put(16.00,4.00){\makebox(0,0)[cc]{$x_{l}$}}
\put(34.00,4.00){\makebox(0,0)[cc]{$x_{i}$}}
\put(41.00,4.00){\makebox(0,0)[cc]{$x_{1}$}}
\put(53.00,17.00){\makebox(0,0)[cc]{$_{G_{n-l+i-1}}$}}
\put(29.07,21.87){\line(-1,0){12.00}}
\put(27.00,10.00){\line(0,1){0.07}}
\put(25.17,16.03){\oval(12.33,6.07)[]}
\put(29.00,13.00){\line(0,-1){0.07}}
\put(68.00,17.00){\makebox(0,0)[cc]{$+$}}
\put(79.00,17.00){\makebox(0,0)[cc]{$(i\leftrightarrow l)$}}
\put(35.00,10.00){\line(0,1){14.07}}
\put(33.00,22.00){\line(0,-1){12.07}}
\put(29.00,10.00){\line(0,1){3.07}}
\put(27.00,13.00){\line(0,-1){3.07}}
\end{picture}\end{center}
\caption{Schwinger-Dyson equation for planar graphs in the large N limit}
                                         \label{fig2}
\end{figure}
Now let us consider a modification of
the right hand side of (\ref {pl}) for correlation functions
corresponding to the rainbow diagrams
\begin {equation} 
                                                          \label {rcf}
\lim _{N \to \infty}\frac{1}{N^{1+n/2}}<\tr (M(x_{n})...M(x_{1}))>_{rb}=
W_{n+2}(x_{n+2},...x_{1})
\end   {equation} 
There are  modifications in the term representing the interaction and
also in the Schwinger terms.
Indeed, let us  consider all possible contractions of a given point $x_{m}$
with a vertex $v$ of the rainbow diagrams. We have to distinguish the cases
of odd and even $m$. For the even $m=2l$ on the left of this vertex $v$
(see Fig.\ref {fig3}a) one has subgraphs corresponding to rainbow
diagrams of the correlation function (Fig.\ref {fig3}b)
\begin {equation} 
                                                          \label {w1}
<(M(x_{2l})M(x_{2l-1})...M(x_{1})(V_{int})^{k_{1}})_{rj}>
\end   {equation} 
or
\begin {equation} 
                                                          \label {w2}
<(M^{3}(x_{2l})M(x_{2l-1})...M(x_{1})(V_{int})^{k_{1}})_{rj}>
\end   {equation} 
By using
\begin {equation} 
                                                          \label {sas}
<(M(x_{i})...M(x_{k}))_{jj'}>=\frac{\delta _{jj'}}{N}
<\tr(M(x_{i})...M(x_{k})>
\end   {equation} 
these terms reproduce $W^{k_{1}}_{2l}(x_{2l},x_{2l-1}...x_{1})$  and
$W^{k_{1}}_{2l+2}(x_{2l},x_{2l},x_{2l},x_{2l-1}...x_{1})$,
respectively.
The rest of the diagram  Fig.\ref {fig3}a corresponds to
$W^{k_{2}}_{n-2l+2}(x_{n},...x_{2l+1},x_{2l},x_{2l})$, $k_{2}=k-k_{1}-1$.

One has also to modify contributions from the Schwinger terms, since only
one correlator (in our case the correlator corresponding to $G_{n-l+i-1}$)
can contain the interaction.
\unitlength=1mm
\special{em:linewidth 0.4pt}
\linethickness{0.4pt}
    \begin{figure}
\begin{center}
\begin{picture}(140.00,35.00)
\put(134.00,30.00){\makebox(0,0)[cc]{$a)$}}
\put(1.00,10.00){\line(0,1){30.00}}
\put(1.00,40.00){\line(1,0){99.00}}
\put(100.00,38.00){\line(-1,0){97.00}}
\put(3.00,38.00){\line(0,-1){28.00}}
\put(5.00,10.00){\makebox(0,0)[lc]{$...$}}
\put(9.00,10.00){\line(0,1){25.00}}
\put(9.00,35.00){\line(1,0){91.00}}
\put(100.00,33.00){\line(-1,0){89.00}}
\put(11.00,33.00){\line(0,-1){23.00}}
\put(1.00,10.00){\line(0,-1){2.00}}
\put(112.50,26.50){\oval(25.00,37.00)[]}
\put(11.00,10.00){\line(1,0){3.00}}
\put(14.00,10.00){\line(0,1){20.00}}
\put(14.00,30.00){\line(1,0){49.07}}
\put(61.00,28.00){\line(-1,0){45.00}}
\put(16.00,28.00){\line(0,-1){18.00}}
\put(16.00,10.00){\line(1,0){3.07}}
\put(19.07,10.00){\line(-1,0){0.00}}
\put(19.07,10.00){\line(1,0){0.93}}
\put(20.00,10.00){\line(0,1){14.00}}
\put(20.00,24.00){\line(1,0){20.00}}
\put(40.00,22.00){\line(-1,0){18.00}}
\put(22.00,22.00){\line(0,-1){12.00}}
\put(24.00,10.00){\makebox(0,0)[lc]{$...$}}
\put(30.00,10.00){\line(0,1){9.07}}
\put(30.00,19.07){\line(1,0){10.00}}
\put(40.00,17.00){\line(-1,0){8.00}}
\put(32.00,17.07){\line(0,-1){9.07}}
\put(32.00,8.00){\line(-1,0){30.93}}
\put(48.50,15.50){\oval(17.00,19.00)[]}
\put(43.00,24.00){\line(-1,0){3.00}}
\put(40.00,22.00){\line(1,0){1.07}}
\put(18.00,5.00){\dashbox{1.00}(41.00,21.00)[cc]{}}
\put(78.00,8.00){\line(-5,2){21.00}}
\put(63.00,30.00){\line(5,-6){15.00}}
\put(78.00,10.00){\line(-5,6){15.00}}
\put(63.00,28.00){\line(-1,0){3.00}}
\put(57.00,18.00){\line(5,-2){21.00}}
\put(78.00,8.00){\line(3,1){22.00}}
\put(100.00,18.00){\line(-5,-2){21.00}}
\put(79.00,9.67){\line(6,5){21.00}}
\put(79.00,11.00){\line(-5,6){4.00}}
\put(35.00,13.00){\makebox(0,0)[cc]{$j$}}
\put(78.00,16.00){\makebox(0,0)[cc]{$s$}}
\put(66.00,19.00){\makebox(0,0)[cc]{$i$}}
\put(86.00,7.00){\makebox(0,0)[cc]{$r$}}
\put(76.00,2.00){\makebox(0,0)[cc]{$v$}}
\put(113.00,25.00){\makebox(0,0)[cc]{$(k_{2})$}}
\put(49.00,17.00){\makebox(0,0)[cc]{$(k_{1})$}}
\put(34.00,7.00){\makebox(0,0)[cc]{$_{x_{1}}$}}
\put(2.00,5.00){\makebox(0,0)[cc]{$_{x_{n}}$}}
\put(14.00,5.00){\makebox(0,0)[cc]{$_{x_{2l}}$}}
\put(36.00,27.00){\line(1,4){1.00}}
\put(38.00,31.00){\line(-1,-4){1.00}}
\put(79.00,11.00){\line(6,5){21.04}}
\end{picture}
\begin{picture}(120.07,40.00)
\put(120.00,30.00){\makebox(0,0)[cc]{$b)$}}
\put(80.00,17.00){\makebox(0,0)[cc]{$W^{k_{1}}_{2l}$}}
\put(10.00,10.00){\line(0,1){14.00}}
\put(10.00,24.00){\line(1,0){20.00}}
\put(30.00,22.00){\line(-1,0){18.00}}
\put(12.00,22.00){\line(0,-1){12.00}}
\put(14.00,10.00){\makebox(0,0)[lc]{$...$}}
\put(20.00,10.00){\line(0,1){9.07}}
\put(20.00,19.07){\line(1,0){10.00}}
\put(30.00,17.00){\line(-1,0){8.00}}
\put(22.00,17.07){\line(0,-1){9.07}}
\put(38.50,15.50){\oval(17.00,19.00)[]}
\put(33.00,24.00){\line(-1,0){3.00}}
\put(30.00,22.00){\line(1,0){1.07}}
\put(8.00,5.00){\dashbox{1.00}(41.00,21.00)[cc]{}}
\put(39.00,17.00){\makebox(0,0)[cc]{$(k_{1})$}}
\put(24.00,7.00){\makebox(0,0)[cc]{$_{x_{1}}$}}
\put(22.00,8.00){\line(-1,0){18.93}}
\put(3.07,8.00){\line(0,1){10.00}}
\put(3.00,17.00){\oval(2.00,2.00)[t]}
\put(3.00,20.00){\oval(2.00,2.00)[b]}
\put(3.00,19.00){\line(0,1){15.00}}
\put(3.07,34.00){\line(1,0){52.00}}
\put(55.07,34.00){\line(0,-1){16.93}}
\put(55.07,17.07){\line(-1,0){8.00}}
\put(47.00,19.00){\line(1,0){6.07}}
\put(53.07,19.07){\line(0,1){12.93}}
\put(53.07,32.00){\line(-1,0){48.00}}
\put(5.07,32.00){\line(0,-1){22.00}}
\put(5.07,10.00){\line(1,0){4.93}}
\put(4.00,5.00){\makebox(0,0)[cc]{$_{x_{2l}}$}}
\end{picture}
\begin{picture}(140.00,40.00)
\put(134.00,38.00){\makebox(0,0)[cc]{$c)$}}
\put(120.00,1.00){\makebox(0,0)[cc]{$W^{k_{2}}_{n-2l+2}$}}
\put(1.00,10.00){\line(0,1){30.00}}
\put(1.00,40.00){\line(1,0){99.00}}
\put(100.00,38.00){\line(-1,0){97.00}}
\put(3.00,38.00){\line(0,-1){28.00}}
\put(5.00,10.00){\makebox(0,0)[lc]{$...$}}
\put(9.00,10.00){\line(0,1){25.00}}
\put(9.00,35.00){\line(1,0){91.00}}
\put(100.00,33.00){\line(-1,0){89.00}}
\put(11.00,33.00){\line(0,-1){23.00}}
\put(1.00,10.00){\line(0,-1){2.00}}
\put(1.00,8.00){\line(1,0){35.00}}
\put(112.50,26.50){\oval(25.00,37.00)[]}
\put(35.00,8.00){\oval(2.00,2.00)[r]}
\put(37.50,8.00){\oval(2.00,2.00)[l]}
\put(37.00,8.00){\line(1,0){15.00}}
\put(50.00,10.00){\line(-1,0){39.00}}
\put(28.00,4.00){\makebox(0,0)[cc]{$j$}}
\put(43.00,4.00){\makebox(0,0)[cc]{$r$}}
\put(69.00,4.00){\makebox(0,0)[cc]{$_{x_{2l}}$}}
\put(1.00,4.00){\makebox(0,0)[lc]{$_{x_{2l-1}}$}}
\put(113.00,25.00){\makebox(0,0)[cc]{$(k_{2})$}}
\put(53.00,8.00){\line(-1,0){3.00}}
\put(53.00,8.00){\line(-1,0){7.00}}
\put(100.00,26.00){\line(-2,-1){32.00}}
\put(72.00,10.00){\line(2,1){28.00}}
\put(72.00,10.00){\line(4,1){28.00}}
\put(100.00,15.00){\line(-4,-1){27.00}}
\put(50.00,10.00){\line(1,0){18.00}}
\put(72.00,8.00){\line(-1,0){19.00}}
\put(72.00,8.00){\line(4,1){12.00}}
\end{picture}
\end{center}
\caption{One term in the Schwinger-Dyson
equation for rainbow graphs}
                                         \label{fig3}
\end{figure}
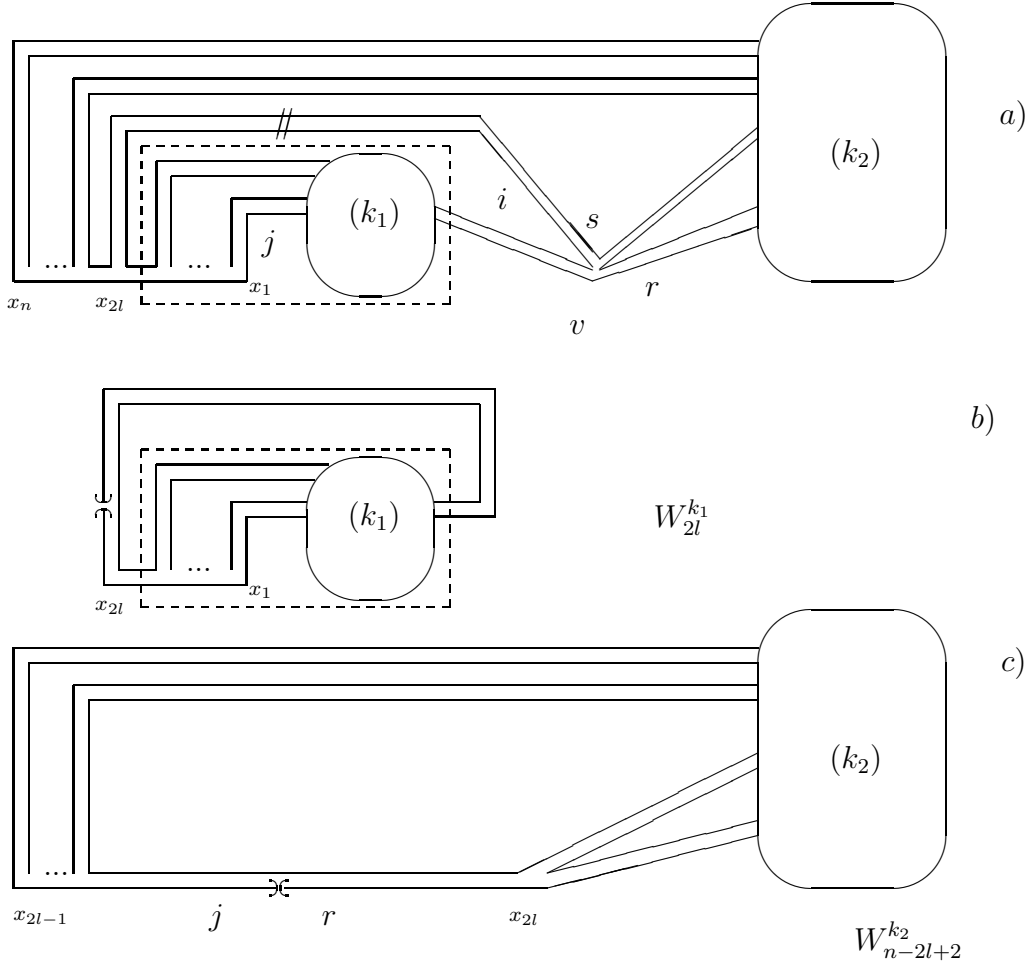
Finally we get the following  system of equations
$$(-\bigtriangleup +m^{2})_{x_{2l}}W_{n}(x_{n},...x_{1})=
g(W_{2l}(x_{2l},x_{2l-1},...x_{1})W_{n-2l+2}(x_{n},...x_{2l+1},x_{2l},x_{2l})
$$
$$
+W_{2l+2}(x_{2l},x_{2l},x_{2l},x_{2l-1},...x_{1})W_{n-2l}(x_{n},...x_{2l+1})
)$$
$$
+\sum _{i< 2l}\delta (x_{2l}-x_{i})W_{n-2l+i-3}(x_{n},...
x_{2l+1},x_{i-1},...x_{1})
W^{(0)}_{2l-i+1}(x_{2l-1},...x_{i+1})$$
\begin {equation} 
                                                          \label {erb}
+\sum _{2l< i}\delta (x_{2l}-x_{i})W_{n+2l-i-3}(x_{n},...x_{i+1},x_{2l-1},...
x_{1})W^{(0)}_{i-2l+1}(x_{i+1},...x_{2l+1});
\end   {equation} 
$$(-\bigtriangleup  +m^{2})_{x_{2l+1}}W_{n}(x_{n},...x_{1})=
g(W_{2l}(x_{2l},x_{2l},...x_{1})W_{n-2l+2}(x_{n},...x_{2l++2},x_{2l+1},
x_{2l+1},x_{2l+1})
$$
$$
+W_{2l+2}(x_{2l+1},x_{2l+1},x_{2l},...x_{1})W_{n-2l}(x_{n},...x_{2l+2}x_{2l-1})
)$$
$$
+\sum _{i< 2l}\delta (x_{2l}-x_{i})W_{n-2l+i-3}(x_{n},...
x_{2l+1},x_{i-1},...x_{1})
W^{(0)}_{2l-i+1}(x_{2l-1},...x_{i+1})$$
\begin {equation} 
                                                          \label {orb}
+\sum _{2l< i}\delta (x_{2l}-x_{i})W_{n+2l-i-3}(x_{n},...x_{i+1},x_{2l-1},
x_{1})W^{(0)}_{i-2l+1}(x_{i+1},...x_{2l+1})
\end   {equation} 

Note that for planar correlation functions it is enough to write down
the equation
differentiated only on $x_{1}$ since the function $G(x_{n},...x_{1})$
is invariant under cyclic permutation.
The rainbow correlation functions loss this property
and we have to write differential equations  for all points $x_{i}$.
In the the right hand side of (\ref {erb}) and (\ref {orb}) enters t
he free rainbow
correlation functions $W^{0}_{n}(x_{n},...x_{1})$. They satisfy the
system of differential equations being the subject of equations (\ref {pl})
with $g=0$.  In distinguish of the interaction case the free rainbow
correlators
posses the property of  invariance under cyclic permutation.
This is due to the fact that free planar
correlators coincide with free rainbow correlators,
i.e. $G^{(0)}_{n}=W^{(0)}_{n}$.
To find $W^{(0)}_{2m}$ it is enough to apply the Wick theorem as it is shown
on Fig.\ref {fig4}a.
\section{The Master Field for Rainbow Diagrams}
\subsection{Free Field Theory}
\setcounter{equation}{0}
Let us consider an algebra generated by operators $A(p)$ and $A^{+}(p)$
satisfying the relations
\begin {equation} 
                                                          \label {fa}
A(p)A^{+}(q)=\delta^{(D)} (p-q).
\end   {equation} 
One can realized this algebra in a space which is an analogue of the
usual Fock space  \cite {Haan,GG}. This space is
generated by the vacuum
vector $|0>$, $A(p)|0>=0$ and n-particle states of n non-identical particles,
\begin {equation} 
                                                          \label {ff}
|p_{1},...p_{n}>=A^{+}(p_{1})...A^{+}(p_{n})|0>
\end   {equation} 
There is no symmetization or antisymmetrization as in the Bose or Fermi cases.
We shall call this Fock space the Boltzmannian Fock space
(it is also called the free Fock space).
One defines
\begin {equation} 
                                                          \label {phi}
\phi (x)=\phi ^{+}(x)+\phi ^{-}(x)=
\frac{1}{(2\pi)^{D/2}}\int \frac{d^{D}p}{\sqrt {p^{2}+m^{2}}}
(A^{+}(p)e^{ipx}+A(p)e^{-ipx})
\end   {equation} 
and therefore
\begin {equation} 
                                                          \label {fpr'}
<0|\phi (x)\phi (y)|0>=D(x-y)=
\frac{1}{(2\pi)^{D}}\int \frac{d^{D}p}{p^{2}+m^{2}}e^{ip(x-y)}
\end   {equation} 
To calculate the n-point correlation function one has to apply a
Boltzmannian Fock space
analog of
the ordinary Wick theorem. The specific feature of
the Wick theorem in this case
is that for a given diagram
one has not additional symmetry factors
related with that an annihilation operator can
be contracted with any creation operator on the right. In the
Boltzmannian Fock space an
annihilation operator can been
contracted only with a nearest creation operator
on the right.  Therefore one sees immediately from the Fig \ref {fig4}b
that the correlation function
\begin {equation} 
                                                          \label {nf}
<0|\phi (x_{2m})...\phi (x_{1})|0>= F^{(0)}_{2m}(x_{2m},...x_{1})
\end   {equation} 
satisfies to the same equations as  $W^{0}_{2m}$, and therefore
$F^{(0)}_{2m}(x_{2m},...x_{1})=W^{(0)}_{2m}(x_{2m},...x_{1})$, i.e.
\begin {equation} 
                                                          \label {fffr}
\lim _{N\to\infty} \frac{1}{N^{1+m}}<\tr (M(x_{2m})...M(x_{1}))>^{(0)}=
<0|\phi (x_{2m})...\phi (x_{1})|0>
\end   {equation} 
\unitlength=1mm
\special{em:linewidth 0.4pt}
\linethickness{0.4pt}
    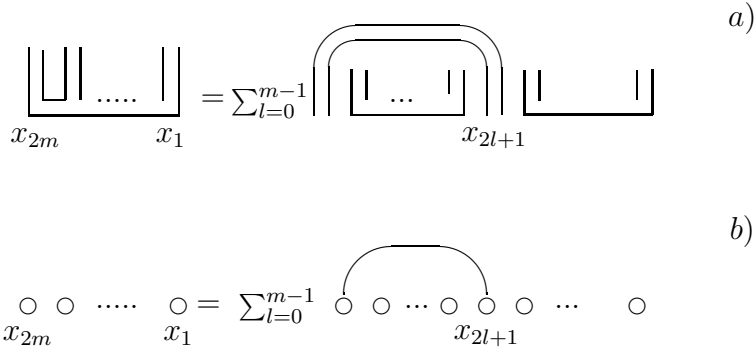
\begin{figure}
\begin{center}
\unitlength=1.00mm
\special{em:linewidth 0.4pt}
\linethickness{0.4pt}
\begin{picture}(125.00,20.00)
\put(5.07,4.93){\line(0,1){9.07}}
\put(7.00,7.00){\line(0,1){0.07}}
\put(7.00,9.00){\line(0,-1){0.07}}
\put(7.07,7.07){\line(1,0){2.93}}
\put(10.00,7.07){\line(0,1){6.93}}
\put(12.00,14.00){\line(0,-1){6.93}}
\put(5.00,5.00){\line(1,0){20.07}}
\put(25.07,5.07){\line(0,1){8.93}}
\put(23.00,14.00){\line(0,-1){0.07}}
\put(14.00,7.00){\makebox(0,0)[lc]{$.....$}}
\put(28.00,7.00){\makebox(0,0)[lc]{$=$}}
\put(32.00,7.00){\makebox(0,0)[lc]{$\sum _{l=0}^{m-1}$}}
\put(48.00,5.00){\line(0,1){6.07}}
\put(50.00,11.00){\line(0,-1){3.93}}
\put(53.00,7.00){\makebox(0,0)[lc]{$...$}}
\put(61.00,11.00){\line(0,-1){3.07}}
\put(63.00,11.00){\line(0,-1){3.07}}
\put(63.07,5.07){\line(-1,0){15.07}}
\put(71.00,11.00){\line(0,-1){0.11}}
\put(71.11,5.11){\line(1,0){16.89}}
\put(88.00,5.11){\line(0,1){6.00}}
\put(86.00,11.00){\line(0,-1){3.89}}
\put(73.00,7.00){\line(0,1){0.11}}
\put(55.50,11.00){\oval(25.00,12.00)[t]}
\put(55.50,11.00){\oval(21.00,8.00)[t]}
\put(68.00,11.00){\line(0,-1){5.89}}
\put(66.00,5.00){\line(0,1){6.11}}
\put(43.00,11.00){\line(0,-1){4.11}}
\put(45.00,5.00){\line(0,1){4.11}}
\put(67.00,2.00){\makebox(0,0)[cc]{$x_{2l+1}$}}
\put(24.00,2.00){\makebox(0,0)[cc]{$x_{1}$}}
\put(6.00,2.00){\makebox(0,0)[cc]{$x_{2m}$}}
\put(100.00,18.00){\makebox(0,0)[cc]{$a)$}}
\put(7.00,7.00){\line(0,1){6.67}}
\put(23.00,7.00){\line(0,1){7.00}}
\put(43.00,5.00){\line(0,1){3.00}}
\put(45.00,8.00){\line(0,1){2.00}}
\put(63.00,5.00){\line(0,1){3.00}}
\put(71.00,5.00){\line(0,1){6.00}}
\put(73.00,7.00){\line(0,1){4.00}}
\put(45.00,9.00){\line(0,1){2.00}}
\end{picture}
\begin{picture}(125.00,25.00)
\put(100.00,15.00){\makebox(0,0)[cc]{$b)$}}
\put(5.00,5.00){\circle{2.00}}
\put(10.00,5.00){\circle{2.00}}
\put(14.00,5.00){\makebox(0,0)[lc]{$.....$}}
\put(25.00,5.00){\circle{2.00}}
\put(25.00,1.00){\makebox(0,0)[cc]{$x_{1}$}}
\put(5.00,1.00){\makebox(0,0)[cc]{$x_{2m}$}}
\put(29.00,5.00){\makebox(0,0)[cc]{$=$}}
\put(33.00,5.00){\makebox(0,0)[lc]{$\sum ^{m-1}_{l=0}$}}
\put(47.00,5.00){\circle{2.00}}
\put(52.00,5.00){\circle{2.00}}
\put(55.00,5.00){\makebox(0,0)[lc]{$...$}}
\put(61.00,5.00){\circle{2.00}}
\put(66.00,5.00){\circle{2.00}}
\put(56.50,6.50){\oval(19.00,13.00)[t]}
\put(71.00,5.00){\circle{2.00}}
\put(86.00,5.00){\circle{2.00}}
\put(75.00,5.00){\makebox(0,0)[lc]{$...$}}
\put(66.00,1.00){\makebox(0,0)[cc]{$x_{2l+1}$}}
\end{picture}

\end{center}
\caption{a) Wick theorem for rainbow graphs in free matrix theory;
b) Wick theorem in the Bolzmannian Fock space}
                                         \label{fig4}
\end{figure}

Let us make a few comments about an operator
realization of the algebra (\ref {pcr})
with $\pi$ satisfying the requirement
\begin {equation} 
                                                          \label {0pr}
\pi (x)|0>=\frac{i}{2}(-\bigtriangleup  +m^{2})_{x}\phi (x)|0>.
\end   {equation} 

 First of all note that the
operator algebra (\ref {pcr}) is not an unique algebra which follows from the
free planar Schwinger-Dyson equation.
Indeed, one gets the same equation from the operator
relation
\begin {equation} 
                                                          \label {mmme}
(-\bigtriangleup  +m^{2})_{x}\phi (x)=\pi (x),~~
[\pi (x),\phi (y)] =-i\delta ^{D}(x-y)|0><0| +K(x,y)
\end   {equation} 
with $K(x,y)$ being the  subject of relations
\begin {equation} 
                                                          \label {kkr}
<0|K(y,x_{1})\phi (x_{2})...\phi (x_{n})|0>
+<0|\phi (x_{1})K(y,x_{2})...\phi (x_{n})|0>+...
\end   {equation} 
$$+<0|\phi (x_{1})...\phi (x_{n-1})K(y,x_{n})|0>=0$$
The simplest solution of (\ref {mmme}) and (\ref {kkr}) is given by
\begin {equation} 
                                                          \label {mpi}
\pi (y)=\frac{i}{2}(-\bigtriangleup +m^{2})_{y}[\phi ^{+}(y)|0><0|
-|0><0|\phi ^{-}(y)],
\end   {equation} 
\begin {equation} 
                                                          \label {sk}
K(y,x)=\frac{i}{2}(-\bigtriangleup +m^{2})_{y}[-\phi ^{+}(x)\phi ^{+}(y)|0><0|-
|0><0|\phi ^{-}(y)\phi ^{-}(x)
\end   {equation} 
$$+\phi ^{+}(y)|0><0|\phi ^{-}(x)
+\phi ^{+}(x)|0><0|\phi ^{-}(y)
]
$$
This gives a hint to write a following operator realization
of the commutation relations (\ref {pcr})
with $\pi $ satisfying the requirement (\ref {0pr})
\begin {equation} 
                                                          \label {pi}
\pi (y)=\frac{i}{2}(-\bigtriangleup +m^{2})_{y}\{\phi ^{+}(y)|0><0|
-|0><0|\phi ^{-}(y)+
\end   {equation} 
$$\sum_{n=1}^{\infty}\int dz_{1} (-\bigtriangleup +m^{2})_{z_{1}}
...\int dz_{n} (-\bigtriangleup +m^{2})_{z_{n}}
[\phi ^{+}(z_{1})...\phi ^{+}(z_{n})\phi ^{+}(y)|0>
<0|\phi ^{-}(z_{n})...\phi ^{-}(z_{1}
$$
$$-\phi ^{+}(z_{1})...\phi ^{+}(z_{n})|0>
<0|\phi ^{-}(y)\phi ^{+}(z_{n})...\phi ^{+}(z_{1})]\}$$

For completeness let us present a  known solution of equation (\ref {pl})
for $D=0$ and $g=0$, i.e equations
\begin {equation} 
                                                          \label {D=0}
\langle \Phi^{2m}\rangle =\sum_{l=0}^{m-1}\langle \Phi^{2l}\rangle
   \langle \Phi^{2m-2l-2}\rangle
\end   {equation} 
where $\Phi=a^{+}+a$, $aa^{+}=1$. Let us denote
$
c_n= \langle \Phi^{2n}\rangle ,\quad c_0=1.
$
Consider the generating function
\begin {equation} 
                                                          \label {gf}
Z(g)=c_0 +c_1 g +...+c_n g^{n}+...= \langle \frac{1}{1-g\Phi^{2}}\rangle
\end   {equation} 
One has
$$
Z(g)^2=c_0^2 +(c_0 c_1 + c_1 c_0)g +...+(c_0 c_{n}
+...+c_{n}c_0) g^{n}+...
$$
 From (\ref {D=0}) one gets
$$
Z(g)^2=c_1 +c_2 g +...+c_{n+1}g^{n}+...
$$
Therefore $Z(g)$ satisfies the equation
$$
gZ(g)^2=Z(g) -1.
$$
One has to take the following solution of this equation
$$
Z(g)={1-\sqrt {1-4g}\over 2g}=1+C_2^1 g+...+{1\over {n+1}}
C_{2n}^n g^{n}+...,
$$
from which we get
$$
c_n=< \Phi^{2n}> =\frac{1}{n+1}C_{2n}^n =\frac{2n!}{n!(n+1)!}
$$

\subsection{Interacting Theory}

We want to derive the Schwinger-Dyson equations for theory
with interaction in the Boltzmannian Fock space. To find
the form of interaction let us consider the following
correlation functions
\begin {equation} 
                                                          \label {icf}
F^{(k)}_{m}(x_{m},...x_{1})=
<0|\phi (x_{m})...\phi (x_{1})(\int dy_{1}:(\phi (y_{1}))^{4}:)
...(\int dy_{k}:(\phi (y_{k}))^{4}:)|0>^{\prime}
\end   {equation} 
where $\phi (x)$ is the free field (\ref {phi})
and $\prime$ means that we do not take into account the diagrams with vacuum
subgraphs. We draw all operators $\phi$ on the staight line.
The operators $\phi (x_{i})$ corresponding to the
external lines are represented by the circles
and the operators $\phi (y_{i})$ corresponding
to the interaction vertices are represented
by the filled circles on Fig.\ref {fig5}.
\unitlength=1mm
\special{em:linewidth 0.4pt}
\linethickness{0.4pt}
    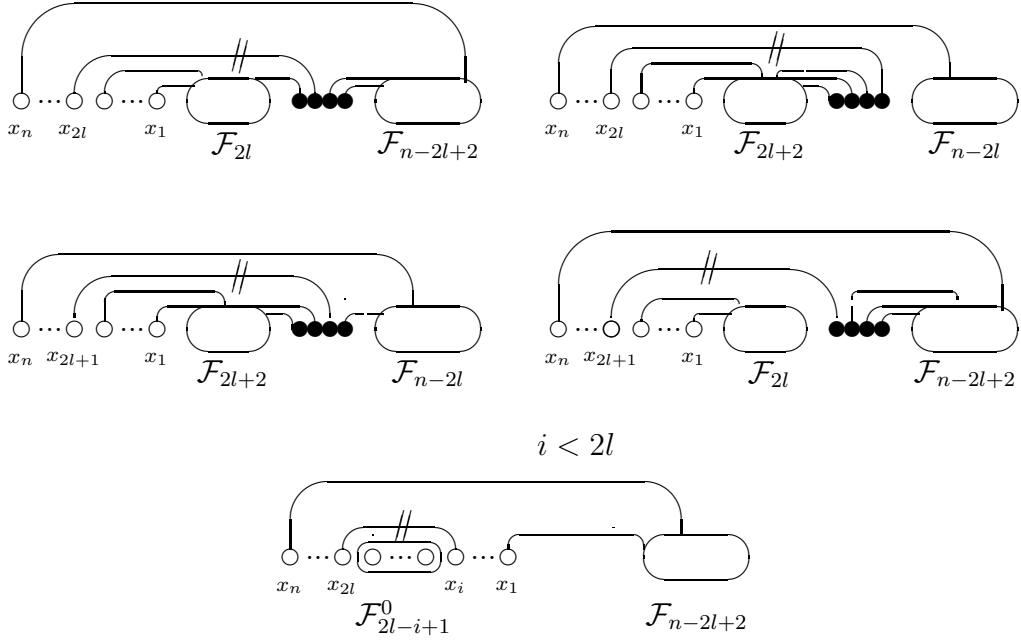
\begin{figure}
\begin{center}
\unitlength=1.00mm
\special{em:linewidth 0.4pt}
\linethickness{0.4pt}
\begin{picture}(70.00,30.00)
\put(2.00,11.00){\line(0,1){2.00}}
\put(2.00,10.00){\circle{2.00}}
\put(4.00,10.00){\makebox(0,0)[lc]{$...$}}
\put(9.00,10.00){\circle{2.00}}
\put(13.00,10.00){\circle{2.00}}
\put(15.00,10.00){\makebox(0,0)[lc]{$...$}}
\put(20.00,10.00){\circle{2.00}}
\put(29.50,10.00){\oval(11.00,6.00)[]}
\put(39.00,10.00){\circle*{2.00}}
\put(41.00,10.00){\circle*{2.00}}
\put(43.00,10.00){\circle*{2.00}}
\put(45.00,10.00){\circle*{2.00}}
\put(56.00,10.00){\oval(14.00,6.00)[]}
\put(2.00,6.00){\makebox(0,0)[cc]{$_{x_{n}}$}}
\put(9.00,6.00){\makebox(0,0)[cc]{$_{x_{2l}}$}}
\put(20.00,6.00){\makebox(0,0)[cc]{$_{x_{1}}$}}
\put(30.00,4.00){\makebox(0,0)[cc]{${\cal F}_{2l}$}}
\put(56.00,4.00){\makebox(0,0)[cc]{${\cal F}_{n-2l+2}$}}
\put(22.50,11.00){\oval(5.00,2.00)[lt]}
\put(25.00,12.00){\line(-1,0){3.00}}
\put(20.50,11.00){\oval(15.00,6.00)[lt]}
\put(19.50,13.50){\oval(13.00,1.00)[rt]}
\put(25.00,11.00){\oval(32.00,10.00)[t]}
\put(31.50,12.50){\oval(59.00,21.00)[t]}
\put(47.50,10.50){\oval(5.00,3.00)[lt]}
\put(50.00,12.00){\line(-1,0){2.93}}
\put(47.50,10.50){\oval(9.00,5.00)[lt]}
\put(52.00,13.00){\line(-1,0){6.00}}
\put(35.00,11.00){\oval(8.00,4.00)[rt]}
\put(35.00,13.00){\line(-1,0){1.93}}
\put(30.00,14.00){\line(1,4){1.07}}
\put(32.00,18.00){\line(-1,-4){1.07}}
\end{picture}
\begin{picture}(70.00,30.00)
      \put(9.00,11.00){\line(0,1){3.00}}
      \put(13.00,11.00){\line(0,1){2.00}}
\put(23.00,13.00){\line(1,0){4.00}}
\put(35.00,14.00){\line(1,0){3.00}}
\put(45.00,14.00){\line(0,-1){4.00}}
\put(2.00,10.00){\circle{2.00}}
\put(4.00,10.00){\makebox(0,0)[lc]{$...$}}
\put(9.00,10.00){\circle{2.00}}
\put(13.00,10.00){\circle{2.00}}
\put(15.00,10.00){\makebox(0,0)[lc]{$...$}}
\put(20.00,10.00){\circle{2.00}}
\put(29.50,10.00){\oval(11.00,6.00)[]}
\put(39.00,10.00){\circle*{2.00}}
\put(41.00,10.00){\circle*{2.00}}
\put(43.00,10.00){\circle*{2.00}}
\put(45.00,10.00){\circle*{2.00}}
\put(56.00,10.00){\oval(14.00,6.00)[]}
\put(2.00,6.00){\makebox(0,0)[cc]{$_{x_{n}}$}}
\put(9.00,6.00){\makebox(0,0)[cc]{$_{x_{2l}}$}}
\put(20.00,6.00){\makebox(0,0)[cc]{$_{x_{1}}$}}
\put(30.00,4.00){\makebox(0,0)[cc]{${\cal F}_{2l+2}$}}
\put(56.00,4.00){\makebox(0,0)[cc]{${\cal F}_{n-2l}$}}
\put(34.50,10.00){\oval(7.00,4.00)[rt]}
\put(35.00,10.50){\oval(12.00,5.00)[rt]}
\put(38.00,13.00){\line(-1,0){6.00}}
\put(37.50,11.00){\oval(11.00,6.00)[rt]}
\put(34.50,13.00){\oval(7.00,2.00)[lt]}
\put(23.00,11.00){\oval(6.00,4.00)[lt]}
\put(21.00,13.00){\oval(16.00,4.00)[t]}
\put(13.00,13.00){\line(0,-1){0.07}}
\put(27.00,13.50){\oval(36.00,7.00)[t]}
\put(30.00,15.00){\line(1,4){1.07}}
\put(32.00,19.00){\line(-1,-4){1.07}}
\put(28.00,15.00){\oval(52.00,10.00)[t]}
\put(54.00,15.00){\line(0,-1){1.93}}
\put(2.00,15.00){\line(0,-1){3.93}}
\end{picture}
\begin{picture}(70,30.00)
\put(13.00,11.00){\line(0,1){3.00}}
\put(2.00,10.00){\circle{2.00}}
\put(4.00,10.00){\makebox(0,0)[lc]{$...$}}
\put(9.00,10.00){\circle{2.00}}
\put(13.00,10.00){\circle{2.00}}
\put(15.00,10.00){\makebox(0,0)[lc]{$...$}}
\put(20.00,10.00){\circle{2.00}}
\put(29.50,10.00){\oval(11.00,6.00)[]}
\put(39.00,10.00){\circle*{2.00}}
\put(41.00,10.00){\circle*{2.00}}
\put(43.00,10.00){\circle*{2.00}}
\put(45.00,10.00){\circle*{2.00}}
\put(56.00,10.00){\oval(14.00,6.00)[]}
\put(2.00,6.00){\makebox(0,0)[cc]{$_{x_{n}}$}}
\put(9.00,6.00){\makebox(0,0)[cc]{$_{x_{2l+1}}$}}
\put(20.00,6.00){\makebox(0,0)[cc]{$_{x_{1}}$}}
\put(30.00,4.00){\makebox(0,0)[cc]{${\cal F}_{2l+2}$}}
\put(56.00,4.00){\makebox(0,0)[cc]{${\cal F}_{n-2l}$}}
\put(34.50,10.00){\oval(7.00,4.00)[rt]}
\put(35.00,10.50){\oval(12.00,5.00)[rt]}
\put(38.00,13.00){\line(-1,0){6.00}}
\put(23.00,11.00){\oval(6.00,4.00)[lt]}
\put(21.00,13.00){\oval(16.00,4.00)[t]}
\put(13.00,13.00){\line(0,-1){0.07}}
\put(45.00,14.00){\line(0,-1){0.07}}
\put(30.00,15.00){\line(1,4){1.07}}
\put(32.00,19.00){\line(-1,-4){1.07}}
\put(28.00,15.00){\oval(52.00,10.00)[t]}
\put(54.00,15.00){\line(0,-1){1.93}}
\put(2.00,15.00){\line(0,-1){3.93}}
\put(26.00,11.50){\oval(34.00,11.00)[t]}
\put(47.50,10.50){\oval(5.00,3.00)[lt]}
\put(48.00,12.00){\line(1,0){2.00}}
\put(23.00,13.00){\line(1,0){4.07}}
\put(13.00,13.00){\line(0,-1){0.07}}
\end{picture}
\begin{picture}(70.00,30.00)
\put(2.00,10.00){\circle{2.00}}
\put(2.00,11.00){\line(0,1){2.00}}
\put(41.00,11.00){\line(0,1){2.00}}
\put(9.00,10.00){\circle{2.00}}
\put(4.00,10.00){\makebox(0,0)[lc]{$...$}}
\put(9.00,10.00){\circle{2.00}}
\put(13.00,10.00){\circle{2.00}}
\put(15.00,10.00){\makebox(0,0)[lc]{$...$}}
\put(20.00,10.00){\circle{2.00}}
\put(29.50,10.00){\oval(11.00,6.00)[]}
\put(39.00,10.00){\circle*{2.00}}
\put(41.00,10.00){\circle*{2.00}}
\put(43.00,10.00){\circle*{2.00}}
\put(45.00,10.00){\circle*{2.00}}
\put(56.00,10.00){\oval(14.00,6.00)[]}
\put(2.00,6.00){\makebox(0,0)[cc]{$_{x_{n}}$}}
\put(9.00,6.00){\makebox(0,0)[cc]{$_{x_{2l+1}}$}}
\put(20.00,6.00){\makebox(0,0)[cc]{$_{x_{1}}$}}
\put(30.00,4.00){\makebox(0,0)[cc]{${\cal F}_{2l}$}}
\put(56.00,4.00){\makebox(0,0)[cc]{${\cal F}_{n-2l+2}$}}
\put(22.50,11.00){\oval(5.00,2.00)[lt]}
\put(25.00,12.00){\line(-1,0){3.00}}
\put(20.50,11.00){\oval(15.00,6.00)[lt]}
\put(19.50,13.50){\oval(13.00,1.00)[rt]}
\put(31.50,12.50){\oval(59.00,21.00)[t]}
\put(47.50,10.50){\oval(5.00,3.00)[lt]}
\put(50.00,12.00){\line(-1,0){2.93}}
\put(47.50,10.50){\oval(9.00,5.00)[lt]}
\put(52.00,13.00){\line(-1,0){6.00}}
\put(48.00,14.00){\oval(14.00,2.00)[t]}
\put(55.00,14.00){\line(0,-1){0.07}}
\put(41.00,14.00){\line(0,-1){0.07}}
\put(24.00,11.50){\oval(30.00,13.00)[t]}
\put(21.00,16.00){\line(1,4){1.13}}
\put(23.00,20.00){\line(-1,-4){1.00}}
\end{picture}
\begin{picture}(70.00,30.00)
\put(2.00,10.00){\circle{2.00}}
\put(4.00,10.00){\makebox(0,0)[lc]{$...$}}
\put(9.00,10.00){\circle{2.00}}
\put(13.00,10.00){\circle{2.00}}
\put(15.00,10.00){\makebox(0,0)[lc]{$... $}}
\put(20.00,10.00){\circle{2.00}}
\put(56.00,10.00){\oval(14.00,6.00)[]}
\put(2.00,6.00){\makebox(0,0)[cc]{$_{x_{n}}$}}
\put(9.00,6.00){\makebox(0,0)[cc]{$_{x_{2l}}$}}
\put(24.00,6.00){\makebox(0,0)[cc]{$_{x_{i}}$}}
\put(30.00,4.00){\makebox(0,0)[cc]{}}
\put(56.00,2.00){\makebox(0,0)[cc]{${\cal F}_{n-2l+2}$}}
\put(13.00,13.00){\line(0,-1){0.07}}
\put(45.00,14.00){\line(0,-1){0.07}}
\put(28.00,15.00){\oval(52.00,10.00)[t]}
\put(54.00,15.00){\line(0,-1){1.93}}
\put(2.00,15.00){\line(0,-1){3.93}}
\put(26.00,10.00){\makebox(0,0)[lc]{$... $}}
\put(16.50,10.00){\oval(11.00,4.00)[]}
\put(24.00,10.00){\circle{2.00}}
\put(16.50,11.00){\oval(15.00,6.00)[t]}
\put(31.00,10.00){\circle{2.00}}
\put(40.00,11.00){\oval(18.00,4.00)[t]}
\put(31.00,6.00){\makebox(0,0)[cc]{$_{x_{1}}$}}
                         \put(16.00,13.00){\line(1,3){1.00}}
                          \put(18.00,16.00){\line(-1,-3){1.00}}
\put(40.00,25.00){\makebox(0,0)[cc]{$i<2l$}}
\put(17.00,2.00){\makebox(0,0)[cc]{${\cal F}^{0}_{2l-i+1}$}}
\put(17.00,12.00){\line(1,4){1.07}}
\put(16.00,12.00){\line(1,4){1.07}}
\end{picture}
   \end{center}
  \caption{Some diagrams contributed to the  Schwinger-
  Dyson equations in the Boltzmannian Fock space }\label{fig5}
  \end{figure}

On Fig.\ref {fig5} we draw all possible contractions of a given
external line with a given vertex.
As we have mentioned above there is no here additional factors related with
symmetry of graphs.
Therefore one has not here the standard factor $1/k!$ in the $k$-th order
of perturbation theory.
This remark leads to an important observation  that to get a set
of equations for correlation functions in the Boltzmannian Fock space we
have to consider instead of the usual exponential factor $\exp \{ V_{int}\}$
the rational function
\begin {equation} 
                                                          \label {rf}
\{1- V_{int}\}^{-1},
\end   {equation} 
(compare with $Z(g)$  (\ref {gf})). Therefore we introduce
the following correlation functions
\begin {equation} 
                                                          \label {ng4}
F_{m}(x_{m},...x_{1})=\sum _{k=0}g^{k}F^{(k)}_{m}(x_{m},...x_{1})=
<0|\phi (x_{m})...\phi (x_{1})\frac{1}{1-V_{int}}|0>^{\prime}
\end   {equation} 
On Fig.\ref {fig5} $V_{int}=g\int dy_{1}(\phi (y_{1}))^{4}$.

Examining all possible contractions of the point $x_{2l}$ we see that
the first two  graphs on Fig.\ref {fig5} reproduce  the first two sums on
the right hand side of (\ref {erb}).
In the similar way
one sees that for odd point the graphs reproduce the first two sum
in the the right hand side of (\ref {orb}).

Performing the normal ordering in the expression
$\phi (x_{m})...\phi (x_{1})$ we get contributions  corresponding
to the Schwingers terms in the correlation function
$(-\bigtriangleup  +m^{2})_{x_{2l}}F_{m}(x_{m},...x_{1})
$.
This consideration proves that $F_{m}$  satisfies to the following equations
$$(-\bigtriangleup +m^{2})_{x_{2l}}F_{n}(x_{n},...x_{1})=
g(F_{2l}(x_{2l},x_{2l-1},...x_{1})F_{n-2l+2}(x_{n},...x_{2l+1},x_{2l},x_{2l})
$$
$$
+F_{2l+2}(x_{2l},x_{2l},x_{2l},x_{2l-1},...x_{1})F_{n-2l}(x_{n},...x_{2l+1})
)$$
$$
+\sum _{i< 2l}\delta (x_{2l}-x_{i})F_{n-2l+i-3}(x_{n},...
x_{2l+1},x_{i-1},...x_{1})
F^{(0)}_{2l-i+1}(x_{2l-1},...x_{i+1})$$
\begin {equation} 
                                                          \label {ebf}
+\sum _{2l< i}\delta (x_{2l}-x_{i})F_{n+2l-i-3}(x_{n},...x_{i+1},x_{2l-1},...
x_{1})F^{(0)}_{i-2l+1}(x_{i+1},...x_{2l+1});
\end   {equation} 
and the similar equations for the $x_{2l+1}$.

Comparing (\ref {ebf}) with (\ref {erb}) we see that $F_{m}$
satisfies to the equations for the rainbow diagrams.
Therefore we get
\begin {equation} 
                                                          \label {th}
\lim _{N\to\infty} \frac{1}{N^{1+m/2}}<\tr (M(x_{m})...M(x_{1}))
\exp \{ \frac{g}{4N}\int dy\tr(M (y))^{4}\}
>^{\bf F}_{rb}=
\end   {equation} 
$$<0|\phi (x_{m})...\phi (x_{1})\frac{1}{1-g\int dy
\phi(y)^{4}}|0>^{\bf BF } $$

 Symbol $<.>^{\bf F}$
 denotes the vacuum expectation value in the ordinary Euclidean bosonic
 Fock space
  and $<.>^{\bf BF}$ denotes nonvacuum diagramms
  in the Boltzmannian Fock space.


In conclusion, a model of quantum field theory with interaction
in the Boltzmannian Fock space has been considered.
We have used the new interaction representation with
a rational function of the
interaction Lagrangian instead of
the exponential function in the standard interaction representation.
The Schwinger-Dyson equations
were derived and it was shown that the perturbation expansion
for the model corresponds to the summation of the rainbow diagramms.
The quantum field with this interaction can be interpreted as the master
field for the rainbow diagramms in the largs $N$ limit matrix model.
The construction of the master field is effective in the sense
that it is purely algebraic and doesn't require  the knowledge of
correlation functions of the theory with the interaction.
Another aspects of quantum field theory  in the Boltzmannian
Fock space including the Minkowskian formulation are considered
in  \cite {AAV2}.

$$~$$

{\bf ACKNOWLEDGMENT}
$$~$$
I.A. and I.V. are supported in part  by
International Science Foundation under the grant M1L000.
I.V. thanks Centro  V. Volterra   Universita di Roma Tor Vergata
where this work was started for the
 kind hospitality.
$$~$$

\end{document}